\documentclass[conference]{IEEEtran}
\IEEEoverridecommandlockouts

\usepackage{cite}
\usepackage{amsmath,amssymb,amsfonts}
\usepackage{algorithmic}
\usepackage{graphicx}
\usepackage{textcomp}
\usepackage{xcolor}
\usepackage{subfig}
\usepackage{fancyhdr}
\def\BibTeX{{\rm B\kern-.05em{\sc i\kern-.025em b}\kern-.08em
    T\kern-.1667em\lower.7ex\hbox{E}\kern-.125emX}}
\begin{document}

\title{A Hybrid Soft Haptic Display for Rendering Lump Stiffness in Remote Palpation\\
\thanks{This work was supported in part by the National Science Foundation Grant 2326453. Pijuan Yu and Anzu Kawazoe contributed equally to this work.}
}

\author{\IEEEauthorblockN{Pijuan Yu}
\IEEEauthorblockA{\textit{Department of Mechanical Engineering} \\
\textit{Texas A\&M University}\\
College Station, USA \\
pijuanyu@tamu.edu}
\and
\IEEEauthorblockN{Anzu Kawazoe}
\IEEEauthorblockA{\textit{Department of Mechanical Engineering} \\
\textit{Texas A\&M University}\\
College Station, USA \\
anzutamu@tamu.edu}
\and
\IEEEauthorblockN{Alexis Urquhart}
\IEEEauthorblockA{\textit{Department of Mechanical Engineering} \\
\textit{Texas A\&M University}\\
College Station, USA \\
lexi.urquhart@tamu.edu}
\and
\IEEEauthorblockN{Thomas K. Ferris}
\IEEEauthorblockA{\textit{Industrial \& Systems Engineering} \\
\textit{Texas A\&M University}\\
College Station, USA \\
tferris@tamu.edu}
\and
\IEEEauthorblockN{M. Cynthia Hipwell}
\IEEEauthorblockA{\textit{Department of Mechanical Engineering} \\
\textit{Texas A\&M University}\\
College Station, USA \\
cynthia.hipwell@tamu.edu}
\and
\IEEEauthorblockN{Rebecca F. Friesen}
\IEEEauthorblockA{\textit{Department of Mechanical Engineering} \\
\textit{Texas A\&M University}\\
College Station, USA \\
rfriesen@tamu.edu}
}

\fancypagestyle{copyright}{
    \fancyhf{}
    \renewcommand{\headrulewidth}{0pt}
    \fancyfoot[C]{\footnotesize \copyright 2026 IEEE. Personal use of this material is permitted. Permission from IEEE must be obtained for all other uses, in any current or future media, including reprinting/republishing this material for advertising or promotional purposes, creating new collective works, for resale or redistribution to servers or lists, or reuse of any copyrighted component of this work in other works.}
}

\maketitle
\thispagestyle{copyright}

\begin{abstract}
Remote palpation enables noninvasive tissue examination in telemedicine, yet current tactile displays often lack the fidelity to convey both large-scale forces and fine spatial details. This study introduces a hybrid fingertip display comprising a rigid platform and a $4\times4$ soft pneumatic tactile display (4.93 mm displacement and 1.175 N per single pneumatic chamber) to render a hard lump beneath soft tissue. This study compares three rendering strategies: a Platform-Only baseline that renders the total interaction force; a Hybrid A (Position + Force Feedback) strategy that adds a dynamic, real-time soft spatial cue; and a Hybrid B (Position + Preloaded Stiffness Feedback) strategy that provides a constant, pre-calculated soft spatial cue.

In a 12-participant lump detection study, both hybrid methods dramatically improved accuracy over the Platform-Only baseline (from 50\% to over 95\%). While the Hybrid B was highlighted qualitatively for realism, its event-based averaging is expected to increase interaction latency in real-time operation. This suggests a trade-off between perceived lump realism and real-time responsiveness, such that rendering choices that enhance realism may conflict with those that minimize latency.

\end{abstract}


\section{Introduction}

Haptic rendering technologies have been investigated for decades in medical teleoperation to solve design problems such as remote palpation, as tactile feedback can significantly enhance an operator's diagnostic performance \cite{okamura2004methods, westebring2008haptics, 10816545}. Among the various methods of providing haptic feedback, normal indentation cutaneous force feedback is often recommended as the preferred stimulation method in telesurgical systems \cite{peeters2008design}. Pacchierotti and Prattichizzo \cite{pacchierotti2023cutaneous} categorize the current normal indentation devices for remote palpation into three main types: (i) rigid, moving platforms with rotational joints that convey surface normals to the fingertip, primarily for rendering 3D surfaces and stiffness \cite{pacchierotti2015cutaneous, chinello2019modular, pompilio2024novel}; (ii) pin-array displays that provide spatially distributed force feedback by raising mechanical pins against the skin, often actuated by shape memory alloys \cite{howe2002remote} or servo motors \cite{feller2004effect, hergenhan2014prototype}; and (iii) soft displays that render rich stiffness information by providing distributed normal stresses and varying the contact area with the fingertip \cite{bicchi2002haptic, kimura2009development}.

\begin{figure}[t]
\centering
\includegraphics[width=3.3 in]{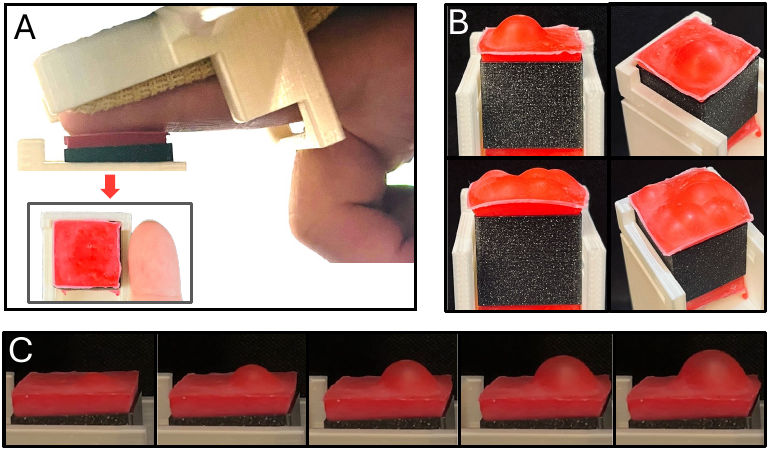}
\caption{\textbf{Hybrid soft tactile display.} (a) A motorized rigid moving platform with a fingertip-size $4\times4$ soft tactile array. (b) Visualization of one bubble (top) and multi bubbles (bottom). (c) Time-lapse of one bubble inflating on the tactile array.}
\vspace{-8pt}
\label{fig_hybrid_haptic_display}
\end{figure}

Each of these approaches, however, presents a distinct set of trade-offs. Rigid moving platforms excel at rendering large-scale, grounded force cues, but as single-point-of-contact devices, they cannot match the high cutaneous sensitivity of the human hand, which requires spatial resolutions approaching the millimeter scale to support spatially-distinguishable tactile sensations (e.g., Merkel disks and Meissner corpuscles at densities of ~500 $cm^{-2}$ and ~100 $cm^{-2}$, respectively) \cite{biswas2019emerging, pacchierotti2023cutaneous, emami2024survey}. Pin-array displays address this by providing spatial feedback, yet their discrete, rigid pins create an unnatural contact interface that makes it hard to replicate the feel of soft tissue palpation in telemedicine \cite{feller2004effect, benko2016normaltouch}. 

Soft actuator technologies, particularly those using materials with a Young's modulus comparable to human tissue (such as silicone rubber), offer significant promise for creating realistic palpation interfaces \cite{levental2007soft}. However, fingertip-scale soft microfluidic arrays are limited by an inherent trade-off between spatial resolution and force output. Existing tactile arrays typically produce less than 0.5 N of blocked force per element \cite{king2008fabrication, haptx_gloves_g1, han2020haptic, shen2023fluid, hartcher2023fingertip, wang2024multiscale, yu2025soft, shan2024multi}. While larger single-chamber actuators can increase this force to around 3 N \cite{kommuri2024fabrication, shao2025wearable, youn2025hapticoil}, this is still insufficient to match the forces applied by clinicians (approx. 6 N) during palpation such as thyroid examinations \cite{chan2025characterization}.

Beyond force limitations, a key challenge in remote palpation and medical simulation is rendering the complex, multilayer mechanical properties of tissue. While rigid \cite{pacchierotti2015cutaneous, chinello2019modular, pompilio2024novel}, pin-based \cite{howe2002remote, feller2004effect, hergenhan2014prototype}, and soft \cite{bicchi2002haptic, kimura2009development} displays have all been employed for lump detection, they face inherent trade-offs in replicating the full diagnostic sensation. In particular, while various haptic softness rendering techniques excel at simulating the compliance of uniform surfaces\cite{fani2017w, tao2021altering, mete2024sori}, they often fail to simulate the specific sensation of a hard inclusion (e.g., a tumor) buried beneath a soft surface within a single, continuous indentation. We hypothesize that this requires a hybrid approach: a rigid platform to render the large-scale force of the surface compliance, combined with a soft actuator to render the fine contact area cues of the underlying hard inclusion. This leads to our first primary research question: to what extent does a hybrid haptic display, which modulates both force and spatial distribution, enhance diagnostic performance compared to a force-only display?

Successfully developing such a device (Fig.~\ref{fig_hybrid_haptic_display}), however, introduces a second critical question regarding the rendering strategy. During pilot testing, preliminary observations suggested that the real-time rendering—updating the soft actuator's pressure based on time-series force data—might produce unrealistic tactile feedback. Specifically, it creates the perception of a bubble growing beneath the fingertip after initial contact has been made. Therefore, an event-based rendering method was designed that provides more stable feedback by averaging the pre-recorded force sensor data over an interaction. But this approach could introduce latency in a remote palpation scenario, as the average force can only be calculated after an interaction is complete. This leads to our second research question: how does the hybrid configuration affect the users' task performance and perceived realism?

To investigate these two research questions, this study conducted a simple lump detection experiment, for now controlling only a vertical stage and a single pneumatic bubble in the center of the array. First, we recorded a palpation exam, in which a user palpated a simulated tissue lump by poking it; no lateral motions were made, and we recorded finger height, applied force to the finger, and video of the exam. Another user then watched the exam video while feeling synchronized haptic feedback constructed from the recorded height and force data, then was asked which exams contained tissue lumps. 
We tested three types of feedback types: 

\noindent
(i) \textbf{Platform-Only (Force Feedback)}: A baseline condition where recorded force is applied solely via vertical stage movement, ignoring the bubble. This mimics many commercially available force feedback devices on the market \cite{WeartTouchDIVER}, and provides a baseline comparison to conditions 2 and 3.

\noindent
(ii) \textbf{Hybrid A (Position + Force Feedback)}: The vertical stage tracks recorded finger height instead of force, while the bubble modulates pressure in real-time to reconstruct force.

\noindent
(iii) \textbf{Hybrid B (Position + Preloaded Stiffness Feedback)}: The stage tracks height, but the bubble is inflated prior to contact, matching the expected stiffness of the soon-to-be-palpated tissue. While this method prevents real-time playback, we hypothesized that it may feel more realistic. The following  section describes the apparatus and rendering strategies.

\section{Materials and Methods}\label{sensing}

To create a modular software architecture capable of low-latency, synchronized visuo-haptic feedback, we used the Robot Operating System 2 (ROS 2) for communication. A unified control loop operates at 100 Hz to manage all high-frequency data streams, including actuator commands, sensor feedback, and user survey data.

\subsection{Hardware}

\begin{figure}[t]
\centering
\includegraphics[width=3.4 in]{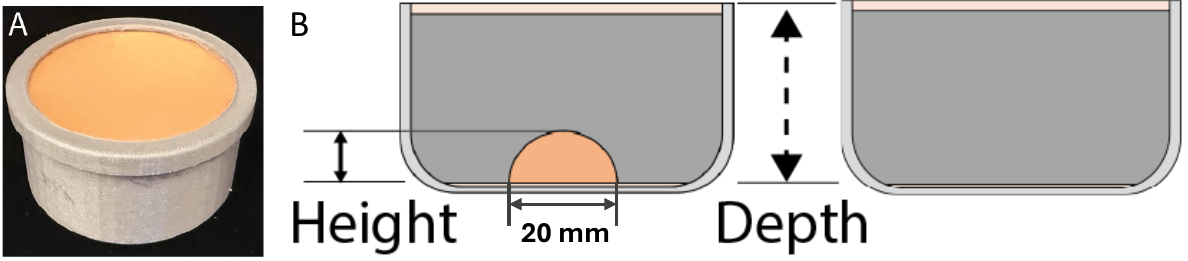}
\caption{\textbf{Tissue phantom.} (a) External view of the tissue phantom. The top surface was covered by a 1~mm skin layer of Ecoflex~00-10, while the interior cavity was filled with loose slime. (b) Diagram of the internal structure, showing a 20 mm diameter, 10 mm height polylactic acid (PLA) inclusion embedded within the 39 mm deep tissue phantom.}
\label{fig_tissuephantom}
\vspace{-5pt}
\end{figure}

\begin{figure*}[t]
\centering
\subfloat[Rigid Platform Position - Force]{\includegraphics[width=2.35in]{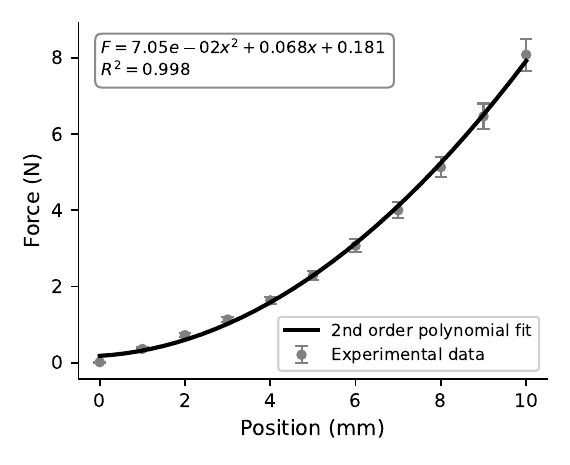}\label{fig:position_char}}
\hfil
\subfloat[Bubble Pressure - Force]{\includegraphics[width=2.35in]{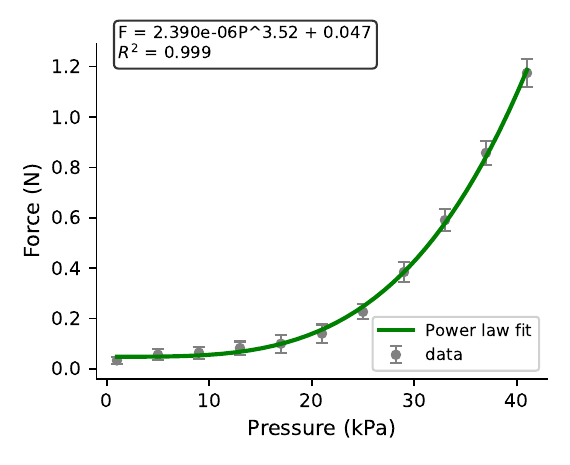}\label{fig:pressure_char}}
\subfloat[Bubble Time - Displacement]{\includegraphics[width=2.35in]{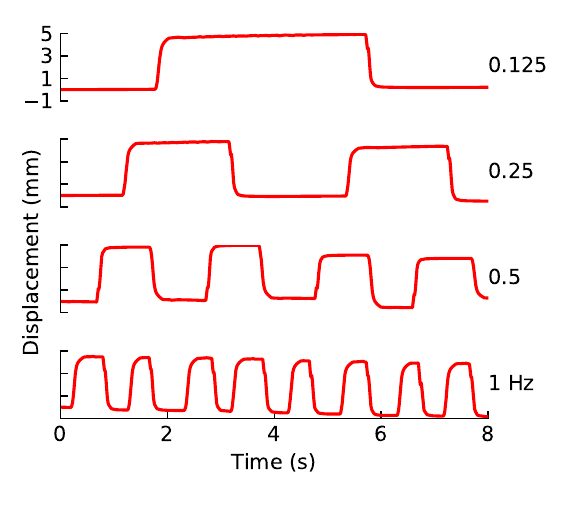}\label{fig:displacement_char}}
\hfil
\caption{\textbf{Haptic display device characterization.} (a) The rigid platform exhibits a nonlinear position-to-force response. (b) A pneumatic bubble shows a nonlinear pressure-to-force relationship. (c) The bubble's maximum vertical displacement is characterized across multiple frequencies.}
\label{fig:characterization}
\vspace{-10pt}
\end{figure*}

\begin{figure}[t]
\centering
\includegraphics[width=3.4 in]{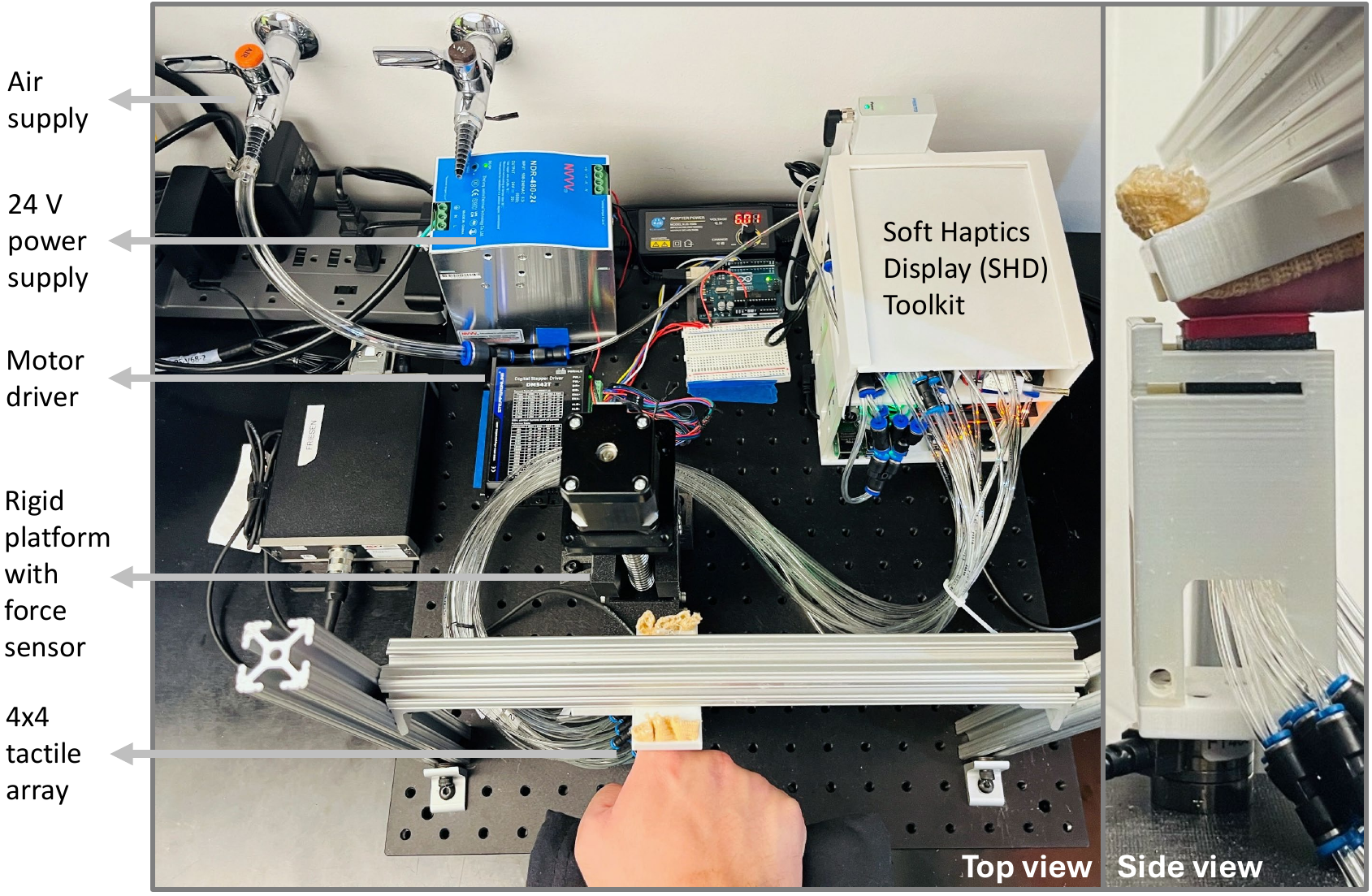}
\caption{\textbf{Haptic display hardware overview.}}
\vspace{-10pt}
\label{fig_hardware_display}
\end{figure}

The haptic display was driven by pre-recorded data captured from a human operator as they performed a single-finger palpation on silicone rubber tissue phantoms, some of which included embedded lumps (Fig.~\ref{fig_tissuephantom}). During this recording, a Tekscan $4\times4$ resistive force-sensing array recorded fingertip pressure distribution, while an OptiTrack motion capture system simultaneously tracked finger position, with all data streams synchronized and sampled at 100 Hz.

The haptic display system architecture is shown in Fig. \ref{fig_hybrid_haptic_display}. The user interface comprises a 3D‑printed finger bracket inclined at 15° to maintain full pad contact with the tactile array. Beneath the fingertip lies a $4 \times 4$ soft pneumatic tactile array comprising 16 hollow, bottomless channels arranged on a 20 mm $\times$ 20 mm grid with 4 mm center-to-center spacing. The actuator is fabricated from two layers of silicone, including a compliant Ecoflex 00-10 top membrane for large deflections and a Dragon Skin 10 Medium substrate for tubing connections. Building upon the initial prototype from the Soft Haptics Display (SHD) Toolkit \cite{yu2025soft}, the current design features two key modifications: the thickness of the top silicone membrane was increased to 1 mm, and the chamber geometry was changed to a 3 mm square.

Each channel is individually actuated by a custom pneumatic control system comprising a proportional pressure regulator and sixteen solenoid valves, allowing for high-speed modulation of tactile patterns. The complete control architecture and communication protocol are detailed in SHD Toolkit \cite{yu2025soft}. To control and measure user interaction, an ATI Nano17 force/torque sensor is integrated directly beneath the actuator. This entire assembly is mounted on a motorized vertical stage (Nema 17 stepper motor), which provides precise positioning of the display relative to the user's finger.

\subsection{Device Characterization}
We characterized the haptic display's output to create accurate control models for both the rigid platform and the pneumatic actuator (see Fig.~\ref{fig:characterization}). The system's response was mapped by relating: (i) vertical displacement of the rigid platform to force, (ii) pneumatic pressure within the bubble to force, and (iii) pneumatic pressure to bubble vertical displacement.

Force characterization (i and ii) was performed with the first author’s index finger in contact with the device to capture the coupled system dynamics (see Fig.~\ref{fig_hardware_display} side view). A consistent pre-loading procedure ensured reliable fingertip contact conditions across all trials. Force, position, and pressure data were sampled at 100 Hz to align with the main control loop. Displacement characterization (iii) was performed using a Laser Doppler Vibrometer without finger contact.

The rigid platform's displacement-force data across 11 position levels (0--10 mm) exhibited a nonlinear relationship that was accurately captured by a second-order polynomial model (Fig.~\ref{fig:position_char}):
\begin{equation}
F = k_{2}x^{2} + k_{1}x + k_{0},
\label{eq:disp_force}
\end{equation}
where $F$ is the rendered force (N) and $x$ is the platform displacement (mm). The polynomial fit yielded an excellent correlation ($R^{2}=0.9983$) across a force span of $0$--$9.58$ N.

The pressure-force relationship for an actuated bubble was best described by a power-law fit (Fig.~\ref{fig:pressure_char}):
\begin{equation}
F = aP^{b} + c_{2},
\label{eq:press_force}
\end{equation}
where $F$ is the rendered force (N) and $P$ is the applied pressure (kPa). The power-law fit demonstrated strong correlation ($R^{2}=0.9987$) across a force span of $0.033$--$1.175$ N.

The vertical displacement of the bubble in the pneumatic actuator was measured by the Laser Doppler Vibrometer and the maximum vertical displacement generated by the bubble is 4.931 mm at 41 kPa (Fig.~\ref{fig:displacement_char}). Notably, the displacement trace shows negative values when the pressure returns to 0 kPa. This behavior is attributed to the dynamic inertial overshoot of the silicone membrane during rapid depressurization. The sudden release of elastic energy causes the membrane to momentarily deflect inward (concave) before settling to its neutral flat state.

\subsection{Rigid Platform Control}

The system renders large-scale contact forces by driving the entire soft actuator as a 1-DoF (degree of freedom) rigid platform.  We implemented two closed-loop control modalities, using Force Mapping for the Platform-Only condition and Position Mapping for the Hybrid conditions.

\subsubsection{Force Mapping for \textbf{Platform-Only}} This modality directly replicates a pre-recorded, one-dimensional time-serial force profile. The summed force from the sensing array serves as the desired setpoint, and a closed-loop Proportional-Derivative (PD) controller adjusts the vertical position of the motorized stage to minimize the error between the desired force and the force measured by the integrated sensor. Recorded position data and bubble actuation are not used in this mode.

\subsubsection{Position Mapping for \textbf{Hybrid A\&B}} To account for differences in mechanical properties between the recording device and the human finger, this modality translates a recorded indentation depth, $d$, into a target platform position, $x_{p}$. Here, $d$ is defined as the distance the finger penetrates the tissue surface: $d = H_{phantom} - z_{finger}$, where $H_{phantom}$ is the phantom height (39 mm) and $z_{finger}$ is the recorded vertical finger position relative to the table surface. The platform coordinate $x_p$ is zeroed at the height where a baseline contact force (0.5 N) is detected.

The mapping uses a model-based strategy that first converts the indentation depth into a contact force using a Hertzian contact model, then maps that force to the required platform position using the inverse of our system's characterized polynomial relationship. This ensures the user perceives a force profile consistent with the compliance of the originally recorded interaction. The governing function is:

\begin{equation}
x_{p} = 
\begin{cases} 
      x_{retract}, & \text{if } d \le 0 \\
      \alpha F_{target}^{2} + \beta F_{target} + \gamma, & \text{if } d > 0
\end{cases}
\label{eq:position_mapping}
\end{equation}
where $F_{target} = \frac{4E^* R^{\frac{1}{2}} d^{\frac{3}{2}}}{3}$ and $\alpha$, $\beta$, $\gamma$ are the coefficients from the inverse polynomial characterization (Eq. \ref{eq:disp_force}), $x_{retract}$ is a predefined negative position ($- 6$ mm) to indicate disengagement, $R$ is the effective radius of the finger (7.5 mm), and $E^*$ is the effective Young's modulus of the finger and tissue phantom interface \cite{opricsan2016experimental}. This function defines the full platform motion, which is attenuated when combined with the soft actuator to ensure users can perceive both the global platform motion and the localized pneumatic patterns. A PD position controller then drives the platform to the attenuated target position.

\subsection{Force Augmentation via Bubble for \textbf{Hybrid A\&B}:}

The hybrid rendering modes enhance stiffness rendering by combining the rigid platform (using the Position Mapping described above) and the pneumatic bubble. The bubble renders the residual force, which contains high-frequency details indicative of features like lumps.
We implemented two bubble control strategies corresponding to two hybrid conditions described in the introduction: 

\noindent
\textbf{Hybrid A}: The system uses a feedforward-feedback controller operating at 100 Hz to continuously modulate the bubble's pressure. The controller's goal is to minimize the error between the residual measured force and the target residual force profile in real-time. 

\begin{figure}[t]
\centering
\includegraphics[width=3.3 in]{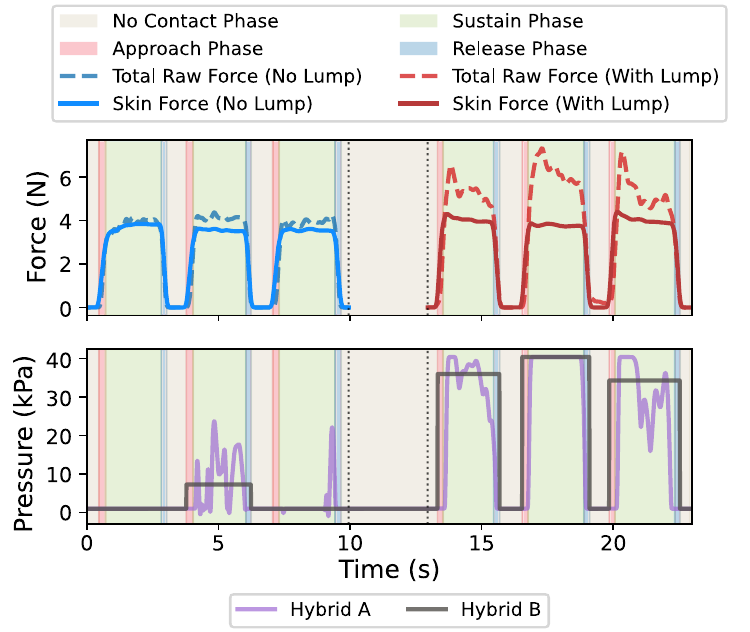}
\caption{\textbf{Hybrid actuation methods.} Top plots show force decomposition; bottom plots show pressure generation during a lump detection task (no-lump: 0-10s; with-lump: 13-23s). The Hybrid A method (purple) tracks residual force at 100 Hz, while the Hybrid B method (dark gray) holds a constant pressure per contact, based on the sustain-phase average.}
\vspace{-10pt}
\label{fig_pressure_control}
\end{figure}

\noindent
\textbf{Hybrid B}: To prevent dynamic inflation artifacts, this strategy uses a phase-based analysis of the interaction. Each poke is segmented into four phases based on the recorded indentation depth $d$ and velocity $\dot{d}$. An indentation depth $d < 0.5$ mm defines the \textit{no contact} phase. During contact, the phase is determined by velocity: \textit{approach} ($\dot{d} < -35$ mm/s), \textit{release} ($\dot{d} > 35$ mm/s), or \textit{sustain}. The average residual force is calculated only during the sustain phase and then mapped to a single, constant pressure command that is applied to the actuator for the duration of the next poke event (see Fig. \ref{fig_pressure_control}).

\subsection{Device Performance Validation}

\begin{figure*}[t]
\centering
\subfloat[Platform-Only: Force Control Performance]{\includegraphics[width=2.35in]{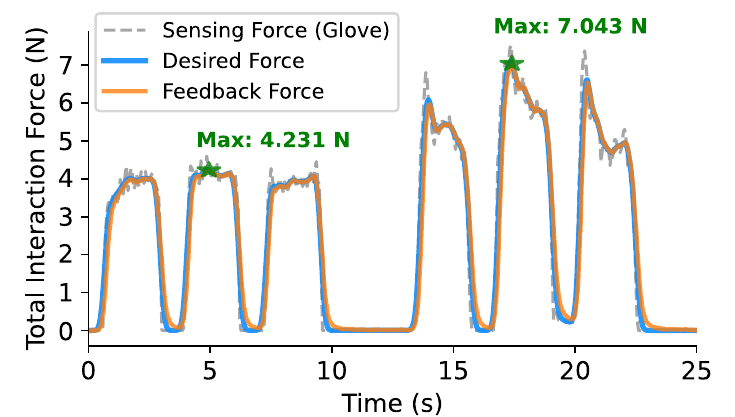}\label{fig:force_control}}
\hfil
\subfloat[Hybrid Modes: Platform Position Control]{\includegraphics[width=2.35in]{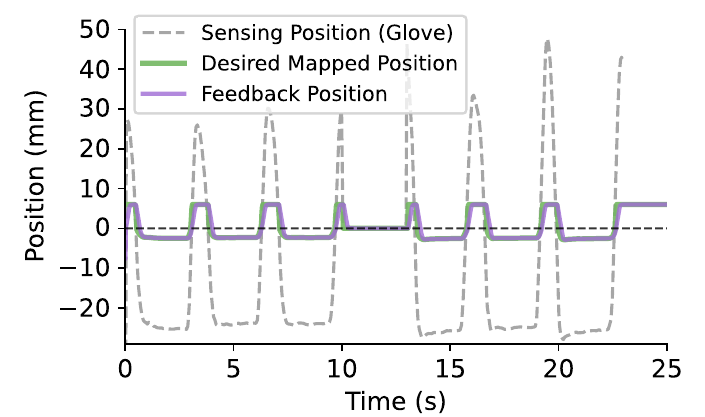}\label{fig:position_control}}
\hfil
\subfloat[Hybrid Modes: Bubble Force Augmentation]{\includegraphics[width=2.35in]{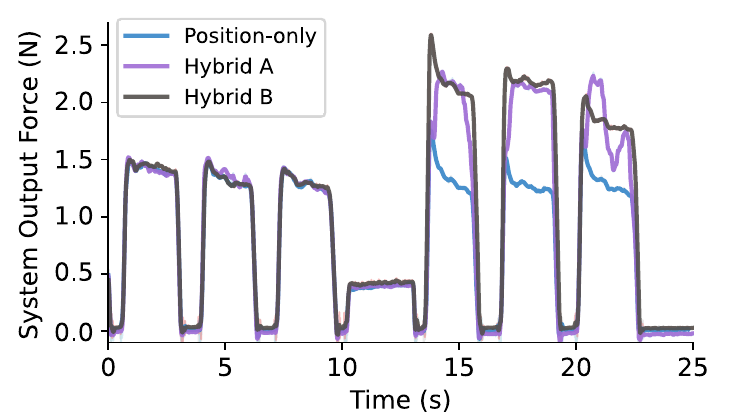}\label{fig:single_bubble_performance}}
\caption{\textbf{System performance validation across the three rendering modes} using a pre-recorded palpation trial (no-lump: 0-10s; with-lump: 13-25s). (a) In \textbf{Platform-Only} mode, the \textbf{force controller} accurately tracks the total interaction force (desired blue, rendered orange). The maximum forces were 4.231 N (no lump) and 7.043 N (with lump), respectively.
(b) In the \textbf{Hybrid} modes, the platform switches to a \textbf{position controller} to track mapped finger position (desired green, rendered purple). 
(c) Comparison of the total measured force from the position-controlled platform without bubble (blue) versus the Hybrid A (purple) and B (dark gray) modes. The gap between the lines demonstrates the force augmentation provided by the bubble.}
\vspace{-10pt}
\label{fig:performance}
\end{figure*}

The haptic display's performance was validated in its three distinct operational modes using pre-recorded palpation data collected by the sensing glove and finger tracking system as the target. In the \textbf{Platform-Only} mode, the rigid platform used a force controller to render the total interaction force. The controller accurately tracked the desired force profile, which reached a maximum of 7.043 N during the trial, with a root-mean-square error (RMSE) of 0.30 N and a correlation coefficient $r$ of 0.99 (Fig.~\ref{fig:force_control}).

In the two \textbf{Hybrid} modes, the platform's role switched to rendering the bulk surface compliance using a position controller. This controller accurately tracked the finger's position with an RMSE of 1.31 mm ($r=0.93$) (Fig.~\ref{fig:position_control}). While tracking position, the platform alone generated a maximum baseline force of 1.79 N. The bubble actuator was then used to augment this force by rendering the high-frequency residual forces (Fig.~\ref{fig:single_bubble_performance}). This augmentation provided an additional 0.91 N (Hybrid A) and 0.97 N (Hybrid B), increasing the total rendered force in the hybrid modes to over 2.60 N. 

Regarding system latency, the rigid platform (stepper motor) exhibits a negligible electromechanical delay (less than 1 ms) compared to the pneumatic system. In contrast, the pneumatic bubble introduces a physical actuation latency of 164.65 $\pm$ 37.42 ms (measured from command to pressure onset\cite{yu2025soft}).

\section{Psychophysical Experiments}\label{experiments}

\begin{figure}[t]
\centering
\includegraphics[width=2.6in]{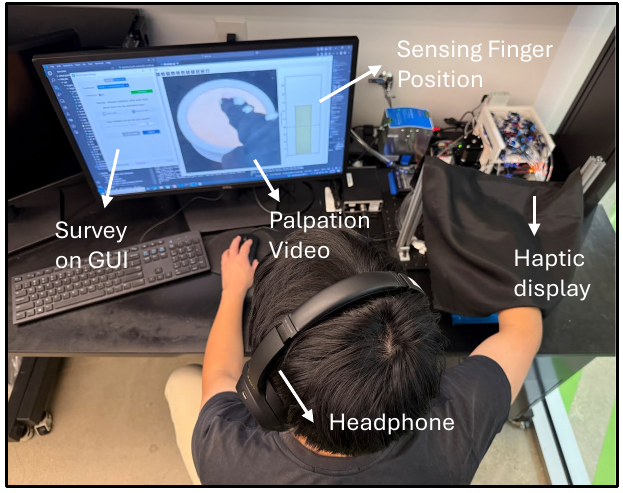}
\caption{\textbf{Psychophysical experiment.} A participant was watching the palpation video with live plot of the sensing finger position and receiving synchronized normal force feedback from haptic display device (covered by a black curtain).}
\label{fig_human_study}
\vspace{-10pt}
\end{figure}

A 12-participant human-subjects study was approved by the Texas A\&M University Institutional Review Board (STUDY2024-0110). Participants (6 male, 6 female; aged 23–31) were healthy, right-handed, and provided informed consent. All participants were engineering students who were novices in clinical palpation; all lacked prior medical training, professional palpation experience, or theoretical knowledge of physical examinations. During the experiment, participants wore noise-canceling headphones playing white noise, viewed pre-recorded videos of a hand palpating a silicone tissue phantom while receiving synchronized haptic feedback (Fig.~\ref{fig_human_study}), and used their multisensory interpretation to make a diagnosis (determine if a lump is present or absent).

\subsection{Stimuli} To ensure repeatable stimuli, a pool of 6 unique 10-second visuo-haptic recordings was created from a human palpating a silicone tissue phantom. Three recordings were of a ``lump present" tissue phantom (20 mm embedded hemisphere) and three were of a ``lump absent" tissue phantom. Each recording featured 3 identical poke motions at the center with a 15-degree inclination, with the operator monitoring indentation depth to ensure consistent contact profiles. Tissue phantom depth and video framing minimized visual lump cues and participants viewed identical camera angles across conditions.

\subsection{Metrics and Procedure}
We employed a within-subjects design with one independent variable, the haptic condition (\textbf{Platform-Only}, \textbf{Hybrid A}, \textbf{Hybrid B}). Dependent metrics were Accuracy (two-alternative forced choice), Confidence (0--100), and Realism (0--100). Participants completed three counterbalanced blocks of the conditions defined above. After training, they performed three trials per block. Each trial presented two sequential stimuli (one with a lump, one without) separated by a 3-second gap (0.5 N preload). Participants identified the lump and rated confidence after each pair. Realism was rated post-experiment.


\subsection{Results} 
\begin{figure}[t]
\centering
\includegraphics[width=3.5in]{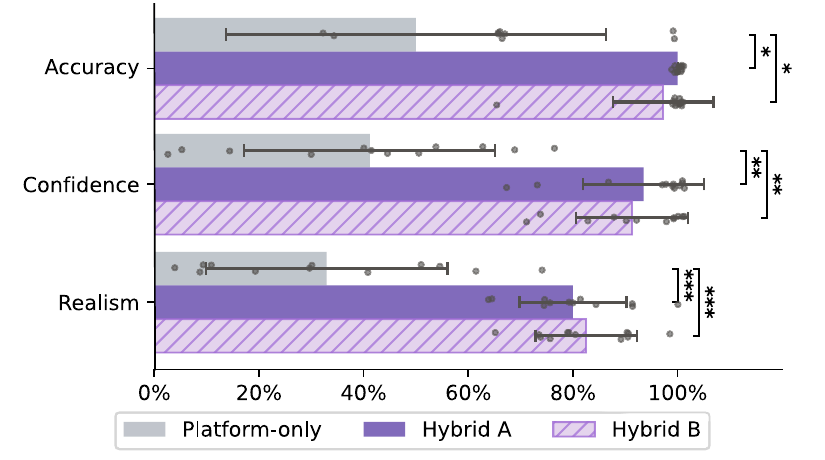}
\caption{\textbf{Result.} Mean accuracy, confidence, and perceived realism for the lump detection study. Error bars denote standard error, and small dots denote individual participants. Statistically significant comparisons between conditions are marked with asterisks; ${*}$, ${**}$ and ${***}$ mean $p < 0.05$, $p < 0.01$ and $p < 0.001$, respectively.}
\label{fig_result}
\end{figure}

\subsubsection{Performance}
Normality assumptions were assessed using Shapiro-Wilk tests. Since the accuracy data violated normality assumptions and employed a within-subjects design, a Friedman test was conducted, revealing a significant effect of haptic feedback condition ($\chi²(2) = 18.2, p < 0.001$) (see Fig.~\ref{fig_result}). Post-hoc pairwise comparisons using Wilcoxon signed-rank tests with Bonferroni correction showed that both the Hybrid A (mean accuracy = $100.0\% \pm 0.0\%$; $p = 0.015$) and Hybrid B (mean accuracy = $97.2\% \pm 9.6\%$; $p = 0.025$) bubble conditions yielded significantly higher accuracy than the Platform-Only condition (mean accuracy = $50.0\% \pm 36.2\%$). There was no significant difference between the two hybrid conditions ($p = 1$).

\subsubsection{Confidence}
Due to violations of normality assumptions in the confidence data, a Friedman test was conducted, revealing a significant main effect of haptic condition ($\chi²(2) = 20.5, p < 0.001$). Post-hoc Wilcoxon signed-rank tests with Bonferroni correction showed that both the Hybrid A (mean confidence = $93.5\% \pm 11.6\%$; $p = 0.001$) and Hybrid B (mean confidence = $91.3\% \pm 10.6\%$; $p = 0.008$) bubble conditions resulted in significantly higher confidence than the Platform-Only condition (mean confidence = $41.2\% \pm 24.0\%$). No significant difference was found between the two hybrid conditions ($p = 0.384$).

\subsubsection{Perceived Realism}
Normality assumptions were confirmed using Shapiro-Wilk tests, allowing for parametric analysis. Following statistical reporting standards in haptics \cite{pacchierotti2015cutaneous}, Mauchly's test was performed to assess the assumption of sphericity. The results indicated that this assumption was violated ($\chi^2(2) = 0.52, p = 0.038$). Consequently, a repeated measures ANOVA with Greenhouse-Geisser correction ($\epsilon = 0.68$) was applied, revealing a significant effect of haptic feedback on perceived realism ($F(1.35, 14.87) = 40.47, p < 0.001$). Post-hoc paired t-tests demonstrated that the Hybrid A (mean = $80.0\% \pm 10.2\%$; $p < 0.001$) and Hybrid B (mean = $82.5\% \pm 9.7\%$; $p < 0.001$) conditions both received significantly higher realism scores than the Platform-Only condition (mean = $32.9\% \pm 23.1\%$). No significant difference in realism was observed between two hybrid conditions.

\section{Discussion}\label{discussion}

A key finding from this experiment is the profound impact of the soft spatially rich tactile display on lump detection, which dramatically improved accuracy despite providing a smaller overall force cue. The \textbf{Platform-Only} condition, with its large force difference between lump and no-lump trials (around 2.8 N shown in Fig.~\ref{fig:force_control}), resulted in chance-level performance. In contrast, the \textbf{Hybrid} conditions succeeded with a much smaller force difference (around 1 N) (Fig.~\ref{fig_result}). Although the bubble actuator does augment the total force, this comparison strongly suggests that a simple increase in force magnitude is insufficient for this task. Instead, human perception appears more sensitive to the complex cues generated by the soft, spatially distributed actuator, which creates a sensation of changing spatial distribution of the force and contact area, consistent with prior haptics research \cite{bicchi2002haptic, fani2017w, tao2021altering, mete2024sori}, than to the pure force magnitude from the rigid platform. This highlights the necessity of soft spatially rich tactile displays to effectively render the detailed features of a lump in remote palpation. 

Our study highlights a trade-off between perceived lump realism and interaction latency, supported by qualitative feedback. The \textbf{Platform-Only} condition was consistently criticized for its lack of shape information; one participant noted, ``The haptic feedback simply felt like my fingertip was being squeezed between two plates". While the \textbf{Hybrid A} method provided this missing spatial cue, it created an unnatural temporal artifact, with a user stating there was a ``delay in feeling the lump on the pad". In contrast, the event-based \textbf{Hybrid B} was often described qualitatively as more realistic by providing a coherent spatial cue and reducing this perceived delay. As one participant commented, ``This one felt close to the real sample because I could feel the bump pressing on my finger to the point that I could almost tell what shape the bump was". However, its across-interaction averaging would increase interaction latency in a real-time remote palpation scenario. These observations suggest that rendering choices that enhance perceived lump realism may conflict with those that minimize latency, indicating an application-dependent balance between spatial realism and real-time responsiveness.

It is important to note that our study did not compare our three display types to additional position-only or bubble-only rendering modes. 
A position-only mode would not render the forces experienced during recorded touch. Since we kept displacement profiles the same between lump and no-lump conditions, without force differences, the two conditions would be indistinguishable.
For a Bubble-Only mode, our pneumatic actuator is designed as a local shape display (augmentation), not a primary force display. Our measured interaction forces ranged from 4 N (no lump) to 7 N (with lump), while the bubble actuator has a maximum output of only 1 N. 

This study has several limitations that present avenues for future work. We intentionally focused on vertical indentation to isolate the psychophysical effects of the control strategies without the confounds of lateral shear or proprioceptive integration found in scanning motions. We only activated a single actuator; future studies will leverage the full $4\times4$ array to render more complex stimuli by varying lump stiffness, size, depth, and mobility. 
While our playback architecture ensures stimulus consistency, the next step is a fully-integrated, real-time remote palpation system that addresses the latency and stability challenges. Furthermore, future studies should involve medical professionals to evaluate the device's clinical utility.




\bibliographystyle{IEEEtran}
\bibliography{references}

@article{okamura2004methods,
  title={Methods for haptic feedback in teleoperated robot-assisted surgery},
  author={Okamura, Allison M},
  journal={Industrial Robot: An International Journal},
  volume={31},
  number={6},
  pages={499--508},
  year={2004},
  publisher={Emerald Group Publishing Limited}
}

@article{westebring2008haptics,
  title={Haptics in minimally invasive surgery--a review},
  author={Westebring--van der Putten, Eleanora P and Goossens, Richard HM and Jakimowicz, Jack J and Dankelman, Jenny},
  journal={Minimally Invasive Therapy \& Allied Technologies},
  volume={17},
  number={1},
  pages={3--16},
  year={2008},
  publisher={Taylor \& Francis}
}

@inproceedings{peeters2008design,
  title={Design considerations for lateral skin stretch and perpendicular indentation displays to be used in minimally invasive surgery},
  author={Peeters, Koen and Sette, Mauro and Goethals, Pauwel and Vander Sloten, Jos and Van Brussel, Hendrik},
  booktitle={International Conference on Human Haptic Sensing and Touch Enabled Computer Applications},
  pages={325--330},
  year={2008},
  organization={Springer}
}

@article{pacchierotti2023cutaneous,
  title={Cutaneous/tactile haptic feedback in robotic teleoperation: Motivation, survey, and perspectives},
  author={Pacchierotti, Claudio and Prattichizzo, Domenico},
  journal={IEEE Transactions on Robotics},
  volume={40},
  pages={978--998},
  year={2023},
  publisher={IEEE}
}

@article{pacchierotti2015cutaneous,
  title={Cutaneous feedback of fingertip deformation and vibration for palpation in robotic surgery},
  author={Pacchierotti, Claudio and Prattichizzo, Domenico and Kuchenbecker, Katherine J},
  journal={IEEE Transactions on Biomedical Engineering},
  volume={63},
  number={2},
  pages={278--287},
  year={2015},
  publisher={IEEE}
}

@article{chinello2019modular,
  title={A modular wearable finger interface for cutaneous and kinesthetic interaction: control and evaluation},
  author={Chinello, Francesco and Malvezzi, Monica and Prattichizzo, Domenico and Pacchierotti, Claudio},
  journal={IEEE Transactions on Industrial Electronics},
  volume={67},
  number={1},
  pages={706--716},
  year={2019},
  publisher={IEEE}
}

@inproceedings{pompilio2024novel,
  title={A novel wearable sensing device enabling remote palpation},
  author={Pompilio, Michele and D’Aurizio, Nicole and Baldi, Tommaso Lisini and Franco, Leonardo and Gabriele, Guido and Prattichizzo, Domenico},
  booktitle={2024 IEEE Haptics Symposium (HAPTICS)},
  pages={149--156},
  year={2024},
  organization={IEEE}
}

@inproceedings{feller2004effect,
  title={The effect of force feedback on remote palpation},
  author={Feller, Ross L and Lau, Camilla KL and Wagner, Christopher R and Perrin, Douglas P and Howe, Robert D},
  booktitle={IEEE International Conference on Robotics and Automation, 2004. Proceedings. ICRA'04. 2004},
  volume={1},
  pages={782--788},
  year={2004},
  organization={IEEE}
}

@inproceedings{hergenhan2014prototype,
  title={Prototype of a haptic display for the evaluation of sensible haptic feedback in remote palpation},
  author={Hergenhan, Jan and Alagi, Hosam and W{\"o}rn, Heinz and Uhl, Michael and Schirren, Rebekka and Reiser, Silvano},
  booktitle={2014 IEEE International Symposium on Medical Measurements and Applications (MeMeA)},
  pages={1--6},
  year={2014},
  organization={IEEE}
}

@article{howe2002remote,
  title={Remote palpation technology},
  author={Howe, Robert D and Peine, William J and Kantarinis, DA and Son, Jae S},
  journal={IEEE Engineering in Medicine and Biology Magazine},
  volume={14},
  number={3},
  pages={318--323},
  year={2002},
  publisher={IEEE}
}

@article{bicchi2002haptic,
  title={Haptic discrimination of softness in teleoperation: the role of the contact area spread rate},
  author={Bicchi, Antonio and Scilingo, Enzo Pasquale and De Rossi, Danilo},
  journal={IEEE Transactions on Robotics and Automation},
  volume={16},
  number={5},
  pages={496--504},
  year={2002},
  publisher={IEEE}
}

@inproceedings{kimura2009development,
  title={Development of a contact width sensor for tactile tele-presentation of softness},
  author={Kimura, Fuminobu and Yamamoto, Akio and Higuchi, Toshiro},
  booktitle={RO-MAN 2009-The 18th IEEE International Symposium on Robot and Human Interactive Communication},
  pages={34--39},
  year={2009},
  organization={IEEE}
}

@article{king2008fabrication,
    author = {King, Chih-Hung and Franco, Miguel and Culjat, Martin O. and Higa, Adrienne T. and Bisley, James W. and Dutson, Erik and Grundfest, Warren S.},
    title = {Fabrication and Characterization of a Balloon Actuator Array for Haptic Feedback in Robotic Surgery},
    journal = {Journal of Medical Devices},
    volume = {2},
    number = {4},
    pages = {041006},
    year = {2008},
    month = {11},
    issn = {1932-6181},
    doi = {10.1115/1.2996593},
    url = {https://doi.org/10.1115/1.2996593},
    eprint = {https://asmedigitalcollection.asme.org/medicaldevices/article-pdf/2/4/041006/5809375/041006_1.pdf},
}

@article{han2020haptic,
  title={Haptic surface display based on miniature dielectric fluid transducers},
  author={Han, Amy Kyungwon and Ji, Sheng and Wang, Dangxiao and Cutkosky, Mark R},
  journal={IEEE Robotics and Automation Letters},
  volume={5},
  number={3},
  pages={4021--4027},
  year={2020},
  publisher={IEEE}
}

@article{biswas2019emerging,
  title={Emerging material technologies for haptics},
  author={Biswas, Shantonu and Visell, Yon},
  journal={Advanced Materials Technologies},
  volume={4},
  number={4},
  pages={1900042},
  year={2019},
  publisher={Wiley Online Library}
}

@article{emami2024survey,
  title={A survey on haptics: Communication, sensing and feedback},
  author={Emami, Melika and Bayat, Amirhossein and Tafazolli, Rahim and Quddus, Atta},
  journal={IEEE Communications Surveys \& Tutorials},
  year={2024},
  publisher={IEEE}
}

@inproceedings{benko2016normaltouch,
  title={Normaltouch and texturetouch: High-fidelity 3d haptic shape rendering on handheld virtual reality controllers},
  author={Benko, Hrvoje and Holz, Christian and Sinclair, Mike and Ofek, Eyal},
  booktitle={Proceedings of the 29th annual symposium on user interface software and technology},
  pages={717--728},
  year={2016}
}

@article{levental2007soft,
  title={Soft biological materials and their impact on cell function},
  author={Levental, Ilya and Georges, Penelope C and Janmey, Paul A},
  journal={Soft Matter},
  volume={3},
  number={3},
  pages={299--306},
  year={2007},
  publisher={Royal Society of Chemistry}
}

@article{chan2025characterization,
  title={Characterization of Medical Neck Palpation to Inform Design of Haptic Palpation Sensors},
  author={Chan, Angela and Kawazoe, Anzu and Kim, Noah and Friesen, Rebecca Fenton and Ferris, Thomas K and Quek, Francis and Hipwell, M Cynthia},
  journal={Sensors},
  volume={25},
  number={7},
  pages={2159},
  year={2025},
  publisher={MDPI}
}

@inproceedings{shen2023fluid,
  title={Fluid reality: High-resolution, untethered haptic gloves using electroosmotic pump arrays},
  author={Shen, Vivian and Rae-Grant, Tucker and Mullenbach, Joe and Harrison, Chris and Shultz, Craig},
  booktitle={Proceedings of the 36th Annual ACM Symposium on User Interface Software and Technology},
  pages={1--20},
  year={2023}
}

@inproceedings{hartcher2023fingertip,
  title={Fingertip wearable high-resolution electrohydraulic interface for multimodal haptics},
  author={Hartcher-O’Brien, Jess and Mehta, Vatsal and Colonnese, Nicholas and Gupta, Aakar and Bruns, Carson J and Agarwal, Priyanshu and others},
  booktitle={2023 IEEE World Haptics Conference (WHC)},
  pages={299--305},
  year={2023},
  organization={IEEE}
}

@article{shan2024multi,
  title={A Multi-Layer Stacked Microfluidic Tactile Display With High Spatial Resolution},
  author={Shan, Boxue and Liu, Congying and Guo, Yuan and Wang, Yiheng and Guo, Weidong and Zhang, Yuru and Wang, Dangxiao},
  journal={IEEE Transactions on Haptics},
  volume={17},
  number={4},
  pages={546--556},
  year={2024},
  publisher={IEEE}
}

@article{wang2024multiscale,
  title={Multiscale haptic interfaces for metaverse},
  author={Wang, Yuanyi and Liang, Jiamin and Yu, Jinke and Shan, Yao and Huang, Xin and Lin, Weikang and Pan, Qiqi and Zhang, Tianlong and Zhang, Zhengyou and Gao, Yongsheng and others},
  journal={Device},
  year={2024},
  publisher={Elsevier}
}

@misc{haptx_gloves_g1,
  author       = {HaptX Inc.},
  title        = {{HaptX Gloves G1}},
  howpublished = {\url{https://haptx.com/gloves-g1/}},
  year         = {2025},
  note         = {Accessed: 2025-02-11}
}

@inproceedings{yu2025soft,
  title={Soft Haptic Display Toolkit: A Low-Cost, Open-Source Approach to High Resolution Tactile Feedback},
  author={Yu, Pijuan and Urquhart, Alexis and Kawazoe, Anzu and Ferris, Thomas K and Hipwell, M Cynthia and Friesen, Rebecca F},
  booktitle={2025 22nd International Conference on Ubiquitous Robots (UR)},
  pages={59--66},
  year={2025},
  organization={IEEE}
}

@article{shao2025wearable,
  title={Wearable Electrohydraulic Actuation For Salient Full-Fingertip Haptic Feedback},
  author={Shao, Yitian and Shagan Shomron, Alona and Javot, Bernard and Keplinger, Christoph and Kuchenbecker, Katherine J},
  journal={Advanced Materials Technologies},
  volume={10},
  number={12},
  pages={2401525},
  year={2025},
  publisher={Wiley Online Library}
}

@inproceedings{youn2025hapticoil,
  title={HaptiCoil: Soft Programmable Buttons with Hydraulically Coupled Haptic Feedback and Sensing},
  author={Youn, Jung-Hwan and Lee, Seung Heon and Shultz, Craig},
  booktitle={Proceedings of the 2025 CHI Conference on Human Factors in Computing Systems},
  pages={1--16},
  year={2025}
}

@inproceedings{opricsan2016experimental,
  title={Experimental determination of the Young's modulus for the fingers with application in prehension systems for small cylindrical objects},
  author={Opri{\c{s}}an, C and C{\^a}rlescu, V and Barnea, A and Prisacaru, Gh and Olaru, DN and Plesu, Gh},
  booktitle={IOP Conference Series: Materials Science and Engineering},
  volume={147},
  number={1},
  pages={012058},
  year={2016},
  organization={IOP Publishing}
}

@article{fani2017w,
  title={W-FYD: A wearable fabric-based display for haptic multi-cue delivery and tactile augmented reality},
  author={Fani, Simone and Ciotti, Simone and Battaglia, Edoardo and Moscatelli, Alessandro and Bianchi, Matteo},
  journal={IEEE transactions on haptics},
  volume={11},
  number={2},
  pages={304--316},
  year={2017},
  publisher={IEEE}
}

@inproceedings{tao2021altering,
  title={Altering perceived softness of real rigid objects by restricting fingerpad deformation},
  author={Tao, Yujie and Teng, Shan-Yuan and Lopes, Pedro},
  booktitle={The 34th Annual ACM Symposium on User Interface Software and Technology},
  pages={985--996},
  year={2021}
}

@article{mete2024sori,
  title={SORI: A softness-rendering interface to unravel the nature of softness perception},
  author={Mete, Mustafa and Jeong, Haewon and Wang, Wei Dawid and Paik, Jamie},
  journal={Proceedings of the National Academy of Sciences},
  volume={121},
  number={13},
  pages={e2314901121},
  year={2024},
  publisher={National Academy of Sciences}
}

@misc{WeartTouchDIVER,
  author = {{Weart}},
  title = {TouchDIVER},
  year = {2025},
  note = {[Online; accessed 22-September-2025]},
  url = {https://weart.it/our-technology/force-feedback/}
}

@ARTICLE{10816545,
  author={Yu, Pijuan and Batteas, Luke C. and Ferris, Thomas K. and Hipwell, M. Cynthia and Quek, Francis and Friesen, Rebecca F.},
  journal={IEEE Transactions on Haptics}, 
  title={Investigating Passive Presentation Paradigms to Approximate Active Haptic Palpation}, 
  year={2025},
  volume={18},
  number={1},
  pages={208-219},
  doi={10.1109/TOH.2024.3523259}}

@inproceedings{kommuri2024fabrication,
  title={Fabrication and characterization of pneumatic unit cell actuators},
  author={Kommuri, Krishna Dheeraj and Van Beek, Femke E and Kuling, Irene A},
  booktitle={Actuators},
  volume={13},
  number={2},
  pages={45},
  year={2024},
  organization={MDPI}
}

\end{document}